\documentstyle[11pt,aaspp4,flushrt, natbib209]{article}
\def\plotfiddle#1#2#3#4#5#6#7{\centering \leavevmode
\vbox to#2{\rule{0pt}{#2}}
\includegraphics{#1}}

\slugcomment{\it In press,\\
 The Astrophysical Journal}
\lefthead{Manning et al.}
\righthead{Primeval Galaxies}

\def\ie{{i.e.,}}
\def\eg{{e.g.,}}

\def\deg{\ifmmode {^{\circ}}\else {$^\circ$}\fi}

\def\secper{\ifmmode \rlap.{^{s}}\else $\rlap{.}{^{s}} $\fi}

\def\kms{\ifmmode {\rm\,km\,s^{-1}}\else
    ${\rm\,km\,s^{-1}}$\fi}
\def\kmsMpc{\ifmmode {\rm\,km\,s^{-1}\,Mpc^{-1}}\else
    ${\rm\,km\,s^{-1}\,Mpc^{-1}}$\fi}
\def\ergAcm2{\ifmmode {\rm\,ergs\,cm^{-2}\,s^{-1}\,{\rm \AA}^{-1}}\else
    ${\rm\,ergs\,cm^{-2}\,s^{-1}\,\AA^{-1}}$\fi}
\def\ergcm2s{\ifmmode {\rm\,ergs\,cm^{-2}\,s^{-1}}\else
    ${\rm\,ergs\,cm^{-2}\,s^{-1}}$\fi}
\def\ergsHz{\ifmmode {\rm\,ergs\,s^{-1}\,Hz^{-1}}\else
    ${\rm\,ergs\,s^{-1}\,Hz^{-1}}$\fi}
\def\ergs{\ifmmode {\rm\,ergs\,s^{-1}}\else
    ${\rm\,ergs\,s^{-1}}$\fi}
\def\ergsA{\ifmmode {\rm\,ergs\,s^{-1}\,\AA^{-1}}\else
    ${\rm\,ergs\,s^{-1}\,\AA^{-1}}$\fi}
\def\ergs{\ifmmode {\rm\,ergs\,s^{-1}}\else
    ${\rm\,ergs\,s^{-1}}$\fi}
\def\WHz{\ifmmode {\rm\,W\,Hz^{-1}}\else
    ${\rm\,W\,Hz^{-1}}$\fi}
\def\Msun{M_\odot}

\def\spose#1{\hbox to 0pt{#1\hss}}
\def\simlt{\mathrel{\spose{\lower 3pt\hbox{$\mathchar"218$}}
     \raise 2.0pt\hbox{$\mathchar"13C$}}}
\def\simgt{\mathrel{\spose{\lower 3pt\hbox{$\mathchar"218$}}
     \raise 2.0pt\hbox{$\mathchar"13E$}}}

\def\hone{\ion{H}{1}}

\def\lya{Ly$\alpha$}

\def\oii{[\ion{O}{2}] $\lambda3727$}
\def\oiipair{[\ion{O}{2}] $\lambda \lambda 3726,3729$}
\def\oiii{[\ion{O}{3}] $\lambda5007$}
\def\oiiipair{[\ion{O}{3}] $\lambda \lambda 4959,5007$}
\def\nv{\ion{N}{5} $\lambda$1240}
\def\civ{\ion{C}{4} $\lambda$1549}
\def\siii{\ion{Si}{2} $\lambda$1260}
\def\alii{\ion{Al}{2} $\lambda$1670}

\begin{document}

\title{A Serendipitous Search for High-Redshift Ly$\alpha$ Emission: \\
Two Primeval Galaxy Candidates at $z \simeq 3$\altaffilmark{1}}

\author{Curtis Manning, Daniel Stern\altaffilmark{2}, Hyron Spinrad,
\& Andrew J. Bunker\altaffilmark{3} }
\affil{Department of Astronomy, University of California at Berkeley \\
Berkeley, CA 94720 \\
{\tt email: (cmanning,dstern,spinrad,bunker)@bigz.berkeley.edu}}

\altaffiltext{1}{Based on observations at the W.M. Keck Observatory,
which is operated as a scientific partnership among the University of
California, the California Institute of Technology, and the National
Aeronautics and Space Administration.  The Observatory was made possible
by the generous financial support of the W.M. Keck Foundation.}

\altaffiltext{2}{Current address:  Jet Propulsion Laboratory,
California Institute of Technology, Mail Stop 169-327, Pasadena, CA
91109; {\tt stern@zwolfkinder.jpl.nasa.gov}}

\altaffiltext{3}{Current address:  Institute of Astronomy, Madingley
Road, Cambridge, CB3 0HA, England; {\tt email: bunker@ast.cam.ac.uk}}

\begin{abstract}

In the course of our ongoing search for serendipitous high-redshift
\lya\ emission in deep archival Keck spectra, we discovered two very
high equivalent width ($W_{\lambda}^{\rm obs} \simgt 450$ \AA, $2
\sigma$) \lya\ emission line candidates at $z \sim 3$ in a moderate
dispersion ($\lambda / \Delta \lambda \simeq 1200$) spectrogram.  Both
lines have low velocity dispersions ($\sigma_v \sim 60\ \kms$) and
deconvolved radii $r \approx 1\ h_{50}^{-1}$ kpc.  We argue that the
lines are \lya, and are powered by stellar ionization.  The surface
density of robust, high equivalent width \lya\ candidates is estimated
to be $\sim 3 \pm 2\ {\rm arcmin}^{-2}$ per unit redshift at $z \simeq
3$, consistent with the estimate of \citet{Cowie:98}.  The \lya\
emission line source characteristics are consistent with the galaxies
undergoing their first burst of star formation, \ie\ with being
primeval.  Source sizes and velocity dispersions are comparable to the
theoretical primeval galaxy model of \citet{Lin:92} based on the
inside-out, self-similar collapse of an isothermal sphere.  In this
model, star formation among field galaxies is a protracted process.

Galaxies are thought to be able to display high equivalent widths for
only the first $\sim$ few $\times 10^7$ yr.  This time is short in
relation to the difference in look back times between $z=3$ and $z=4$,
and implies that a substantial fraction of strong line-emitting
galaxies at $z=3$ were formed at redshifts $z \leq 4$.  We discuss the
significance of high-equivalent width \lya-emitting galaxies in terms
of the emerging picture of the environment, and the specific
characteristics of primeval galaxy formation at high redshift.

\end{abstract}


\clearpage

\section{Introduction}

Early theoretical models of primeval galaxies suggested that their
\lya\ lines would be highly luminous and diffuse.  But searches on
4m-class telescopes confirmed no such emission lines
\citep{Pritchet:94, Thompson:95} and the search for galaxies at large
look-back times turned to other strategies.  One highly successful
technique is color-selection of the Lyman break in galaxies of redshift
$z \simeq 3$ \citep{Steidel:92, Steidel:96a}.  Lyman-break galaxies
(LBGs) are strongly star-forming, and compact relative to local ${\cal
L}^*$ galaxies \citep{Giavalisco:96b}.  However, LBGs are not generally
thought to be primeval, in part because they have lower
\lya\ equivalent widths than expected of primeval galaxies, in part
because interstellar metallic absorption lines (\eg\ \siii, \alii) are
characteristic of most LBG spectra.

The theoretical upper limit equivalent width for \lya\ emission
produced by stellar photoionization is thought to be
$W_{\lambda}^{\rm rest} \sim 100 - 200$ \AA\ \citep{Charlot:93}, and
can be expected to last only a few$\times 10^7$ yr for a constant star
formation rate (SFR).  Several examples of high-redshift, high
equivalent width (non-AGN) \lya\ emission have been reported
\citep[\eg][]{Dey:98, Weymann:98, Hu:99}.  Few secure examples of
isolated, high equivalent widths that are not \lya\ exist in the
literature \citep[\eg][]{Stockton:98, Stern:99b}.  About half of LBGs
have \lya\ in emission \citep{Steidel:96a, Steidel:98, Lowenthal:97},
with rest equivalent widths $\sim 5 - 35$ \AA.  Likewise, about half
of local actively star-forming galaxies have \lya\ in emission, with
rest equivalent widths in the range $W_{\lambda} \simeq 10 - 35$ \AA\
\citep{Giavalisco:96a}, though a few have $W_{\lambda} \simgt 50
\mbox{ \AA}$.

Local studies show that whether \lya\ is seen in emission or
absorption may depend heavily on the kinematics of the surrounding
neutral \hone\ halo \citep{Legrand:97, Kunth:98a} and the chance
geometry of neutral gas and dust \citep{Giavalisco:96a}.  However,
$W_{\lambda}^{\rm rest}$ is shown to be weakly-correlated with
metallicity \citep{Terlevich:93, Giavalisco:96a}.  The selective
absorption of resonantly scattered \lya\ photons, whose path length in
the cloud is much longer than that of photons of modestly different
wavelength, is thought to be the major factor in quenching the \lya\
line.  The partially extincted \lya\ line is characteristically
asymmetric \citep[\eg][]{Dey:98, Kunth:98a, Legrand:97}, presumably
caused by the absorption of photons from the blue side of the line by
dust in an expanding shell, and compounded by the addition of
red-shifted backscattering off the neutral shell expanding away from
the observer \citep[\eg][]{Legrand:97}.  Though dust-free galaxies are
the most likely to have high equivalent width \lya\ emission, it is
clear that a more evolved galaxy can also have high equivalent width
\lya\ emission emerging within specific zones of active star formation
due to an incomplete covering factor.  Such emission may appear
spatially offset from the region of star formation
\citep[\eg][]{Bunker:00}.  Averaged over the whole galaxy, the
continuum supplied by recent star formation with locally quenched
\lya\ emission will reduce the integrated line equivalent width.

Deep spectra taken at the Keck telescopes regularly reveal
serendipitous high-equivalent width, isolated emission lines
\citep[\eg][]{Stern:99b}.  Indeed, the first confirmed galaxy at $z >
5$ was discovered in this manner \citep{Dey:98}.  \citet{Steidel:98}
obtained a serendipitous spectrum of a strong \lya\ line with
extremely faint continuum near the $\langle z \rangle \simeq 3.09$
structure in the SSA22 field.  Strong emission lines can also be
effectively targeted by using narrow band imaging, in conjunction with
broad band imaging.  Recent observations, using narrow band imaging
with a ``strong equivalent width'' ($W_\lambda^{\rm obs} \geq 77$ \AA)
criterion \citep{Cowie:98} and follow-up spectroscopy \citep{Hu:98}
have disclosed a population of galaxies with what is thought to be
\lya\ in emission.  The galaxies tend to have very weak, sometimes
undetectable, continua.  They are also are quite compact, as noted of
the $z\sim 2.4$ \lya\ emitters found in HST searches 
\citep{Pascarelle:96, Pascarelle:98}.  \citet{Cowie:98} estimate a
surface density of 3.6 arcmin$^{-2}$ per unit redshift for flux
densities $j > 2 \times 10^{-17}~\ergcm2s$, corresponding to line
luminosities ${\cal L}_{\rm Ly\alpha} >1.8 \times 10^{42} \ergs$ for
an Einstein-de Sitter Universe with $H_0 = 50 \kmsMpc$ at $z \approx
3.4$.

We are conducting a search for serendipitous emission lines in deep
archival Keck spectra.  Such a serendipitous search is an emission-line
flux-limited survey, and may find emission lines over a wide range of
redshifts.  Narrow-band surveys are sensitive to candidates over
only small redshift ranges, and require follow-up spectroscopy to
discriminate stellar \lya\ lines from metal and AGN emission lines.
The identification of a population of primeval galaxies may provide
important information about the epoch(s) of galaxy formation, including
data relevant to the integrated global star formation rate, and the
luminsity function of galaxies to photometric limits fainter than
that accessible to color-selection surveys.

One drawback faced by all \lya\ emission line searches is possible
confusion with other isolated, high equivalent width lines such as
\oii\ and H${\alpha}$.  Experience has shown that \oii\ is often
mis-identified as \lya.  When continuum is detected, we may use the
Balmer series or the continuum depression to discriminate, and when
absent we may search for another line, such as \nv\ or \oiii.
However, \nv\ is rarely detected in non-AGN, and the search for \oiii\
often fails because these lines would either fall out of the range of
the spectrograph, or in the high-noise infrared part of the spectrum
which, combined with the large intrinsic dispersion in \oii/\oiii\
\citep{Kennicutt:92}, can make an unambiguous verification untenable.
Alternatively, one may use the known tendency of \lya\ lines to
display a P-Cygni profile to identify this resonance line.  However,
asymmetry cannot be a completely necessary criterion for \lya\ emission
lines since some local sources apparently have symmetric, or ``pure''
profiles \citep{Kunth:98b}.  When emission lines are unresolved, and
have undetected continua, the last recourse is the knowledge that
\oii\ rarely has an equivalent width greater than 100 \AA\ \citep[but
see][]{Stern:99b}.  For these reasons, unambiguous \lya\ lines are
rare, and worthy of extra scrutiny.

We serendipitously found two isolated emission lines on a single
slitlet from a mask centered on the SSA22 field, andq determined that
they are in fact \lya\ lines --- examples of galaxies which are more
compact and have higher \lya\ equivalent widths than known LBGs.  We
report on their equivalent widths, intrinsic radii, velocity
dispersions, and surface/volume density in order to place them in the
picture of the evolving Universe emerging from studies of
high-redshift galaxies.  The source targeted by the slitlet in which
we found our emission line galaxies is the color-selected Lyman-break
galaxy SSA22~C17; we analyze it in parallel for comparison. We adopt
$H_0 = 50\, h_{50}\, \kmsMpc, \Omega_0 = 1,$ and $\Lambda = 0$ unless
otherwise stated.  At $z = 3$, for $\Omega = 1 (0.1)$, 1\farcs0
corresponds to $7.3\, (12.3)\, h_{50}^{-1}$ kpc.

\section{Observations and Data Reduction}

We have obtained deep, moderate-dispersion spectra of $z \simeq 3$
LBGs with the aim of making detailed studies of the ages, kinematics,
dust-content, and abundances of the LBG population (Dey et. al., in
preparation).  The data were taken during the years 1997 to 1999 with
the Low Resolution Imaging Spectrometer \citep{Oke:95} at the
Cassegrain focus on the Keck~II telescope.  The camera uses a Tek
$2048^2$ CCD detector with a pixel scale 0\farcs212 pix$^{-1}$.  On UT
1997 September 10 we obtained moderate-dispersion multislit spectra of
LBGs in the SSA22 field, using the 600 lines mm$^{-1}$ grating blazed
at 5000 \AA.  Slitlet widths were 1\farcs25, resulting in a spectral
resolution of $\sim$ 4.4 \AA\ (FWHM) for sources filling the slitlet,
and a resolution $\lambda/\Delta \lambda_{\rm FWHM} \sim 1200$.
Objects not filling the slitlet will have higher spectral resolution.
For a spatially unresolved source -- one blurred only by atmospheric
seeing -- we estimate a resolution of $\sim$ 2.8 \AA\ (FWHM) based on
\lya\ absorption systems in the quasar SSA22~D14 also observed on the
slitmask reported herein.  Four of the integrations totaling 7200$s$
were of excellent quality, with seeing along the spatial axis of
$0\farcs78$.  Additional observations were made on UT 1997 September
12 and UT 1999 June 14.  Here we concentrate on the data of UT 1997
September 10 for which the signal-to-noise ratio ($S/N$) is highest.

The data reductions were performed using the {\tt IRAF} package, and
followed standard slit spectroscopy procedures.  Flat-fielding, sky
subtraction, cosmic ray removal, and aperture extractions of the
slitmask data were facilitated by the home-grown software package {\tt
BOGUS}, created by D.~Stern, A.~J.~Bunker \& S.~A.~Stanford.
Wavelength calibration was performed using a HgNeAr lamp, employing
telluric sky lines to adjust the wavelength zero-point to the data
frames, compensating for any drift in the wavelength coverage.  Flux
calibration was performed using observations of Wolf 1346
\citep{Massey:90}.

%[EDITOR: FIGURE 1 GOES NEAR HERE]

\begin{figure}[ht]

\plotfiddle{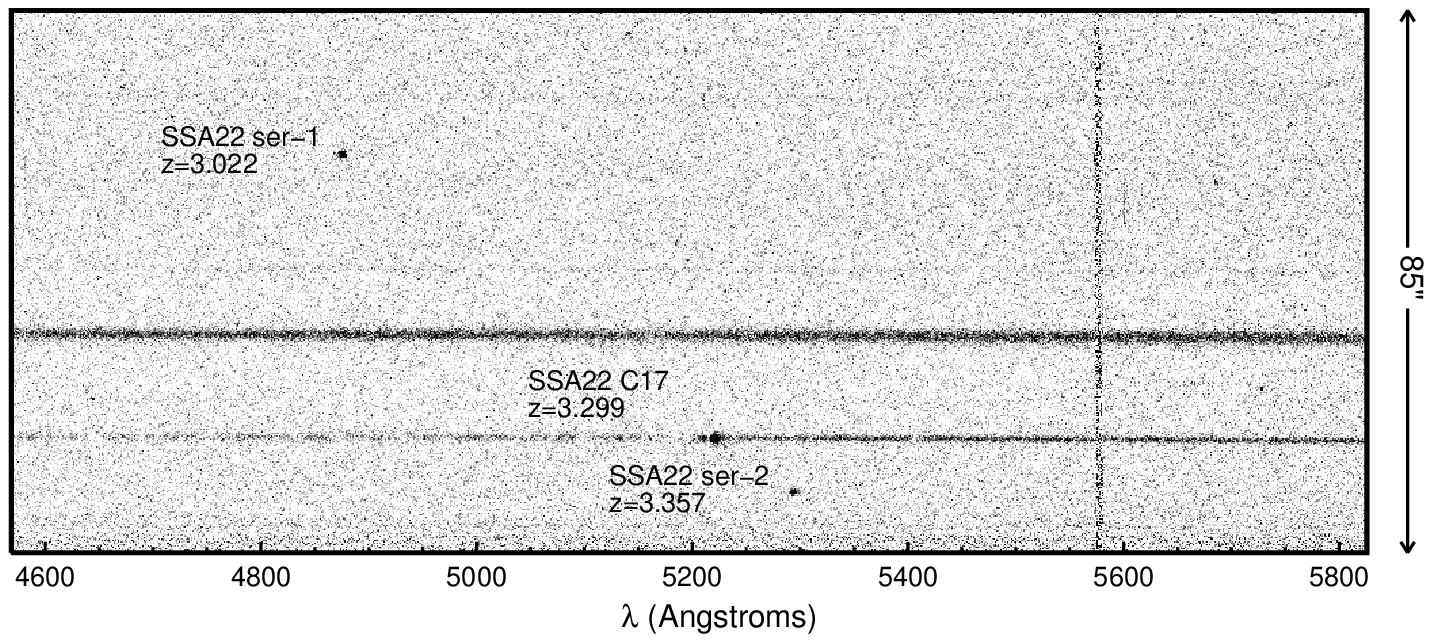}{2.0in}{0}{91}{91}{-190}{-21}

\caption{Two-dimensional spectrogram of the slitlet in the SSA22 field
targeting the LBG SSA22~C17 at $z=3.299$.  Two strong line emitters are
serendipitously identified, ser-1 and ser-2.  The high-equivalent
widths, narrow velocity widths, and lack of secondary spectral features
strongly argue that these are \lya\ emitters at $z \simeq 3$.  Note the
foreground continuum source and the residual of the
[\ion{O}{1}]$\lambda$ 5577 \AA\ skyline.}

\end{figure}

\begin{figure}[ht]
% prg for this in ~/fortran.  find *.f, *.macro, *.eps

\plotfiddle{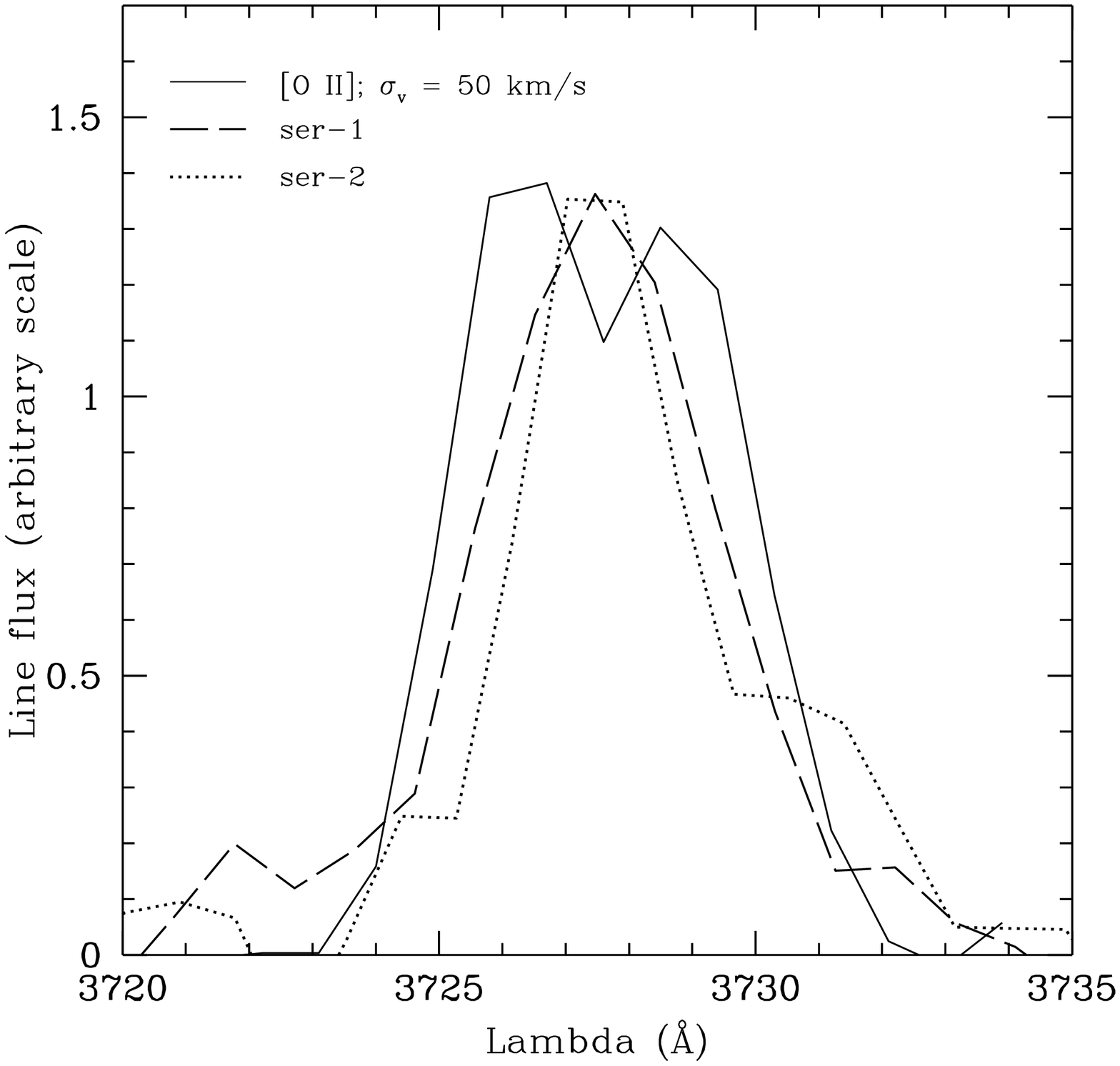}{3.0in}{0}{45}{45}{-146}{-72}

\caption{The \oiipair\ doublet line profile from a source with a 50 \kms\
velocity dispersion, as it would appear in its rest frame, observed
with an effective 2.8 \AA\ resolution of a point-source emitter.
Poisson noise has been added commensurate to that observed in the LRIS
spectra.  The observed emission lines of ser-1 and ser-2 have been
referred to the same wavelength, with their amplitude adjusted to
agree with that of the \oiipair\ lines.  The line profiles of ser-1
and ser-2 are clearly incompatible with an \oiipair\ interpretation.}

\end{figure}

%[EDITOR: FIGURE 2 GOES NEAR HERE]

\section{Spectroscopic Results}

We find two serendipitous, isolated emission lines, ser-1 and ser-2,
in a 100\arcsec\ long slitlet centered on the LBG SSA22~C17 ($z =
3.299$; see Fig.~1).  In lower dispersion spectra, it is often
impossible to distinguish between high equivalent width forms of \oii\
and \lya, unless there is an evident continuum depression, a line
asymmetry, or other distinguishing features.  Though only one of these
two serendipitously discovered sources (ser-2) displays evidence of
the asymmetry characteristic of high-redshift \lya\ emission in LBGs,
we argue that \lya\ is indeed the most likely interpretation for
both. We simulate the \oiipair\ doublet using our 2.8 \AA\ resolution
(for an unresolved source) with 50 \kms\ intrinsic velocity
dispersion, introducing the Poisson noise appropriate to the observed
flux of the median of ser-1 and ser-2 (see Fig.~2).  We conclude that
\oiipair\ is inconsistent with our lines, as the oxygen doublet would
be marginally resolved and has a significantly greater FWHM.  Further,
the absence of associated emission lines argues against H${\beta}$ and
\oiiipair\ interpretations.  Finally, the lines are shortward of 6563
\AA\, so H${\alpha}$ is not a viable identification.  In the
following, we assume what is most certainly the case --- that these
are in fact \lya\ emission lines.  Notably, these spectra display no
detectable continua.

%[EDITOR: TABLE 1 GOES NEAR HERE]

\begin{table}
\caption{Emission Line Properties}
\scriptsize
\begin{center}
\begin{tabular}{cccccccccc}
\tableline
\tableline
Object &
$\lambda^{\rm obs}$  &
$z$ &
$j_{-17}^a$ &
${\cal L}_{42}^b$ &
$W_{\lambda}^{\rm obs}$ &
$W_\lambda^{\rm rest}$ &
$r^c$ &
$\sigma_v^d$ \\
&
(\AA) &
& 
(cgs) &
(cgs) &
(\AA) &
(\AA) &
($h_{50}^{-1}$ kpc) &
(\kms) \\
\tableline
ser-1 & 4889.3 & 3.022 & 1.85 & 1.29 & $\geq 550\, (2\sigma)$ & $\geq 137\,
(2\sigma)$ & $0.8 \pm 0.6$ & $62 \pm 9$ \\
ser-2 & 5296.8 & 3.357 & 1.35 & 1.19 & $\geq 470\, (2\sigma) $ &$\geq 109\,
(2\sigma)$ & $1.1 \pm 0.5$ & $47 \pm 9$ \\
C17 & 5226.4 & 3.299 & 2.60 & 2.20 & 35.6 & 8.3 & 2.8: & 82: \\
\tableline
\end{tabular}
\end{center}
\medskip

\emph{Notes.---} (a) Emission line fluxes are in units of $10^{-17}$
\ergcm2s.  (b) The \lya\ line luminosities are in units of $10^{42}\,
h_{50}^{-2}\, \ergs$.  (c) radius based on deconvolution of source
FWHM.  (d) Deconvolved velocity dispersion  assumes sources fill the
1\farcs25 width of slitlets.

\normalsize
\end{table}

\begin{figure}[ht]

\plotfiddle{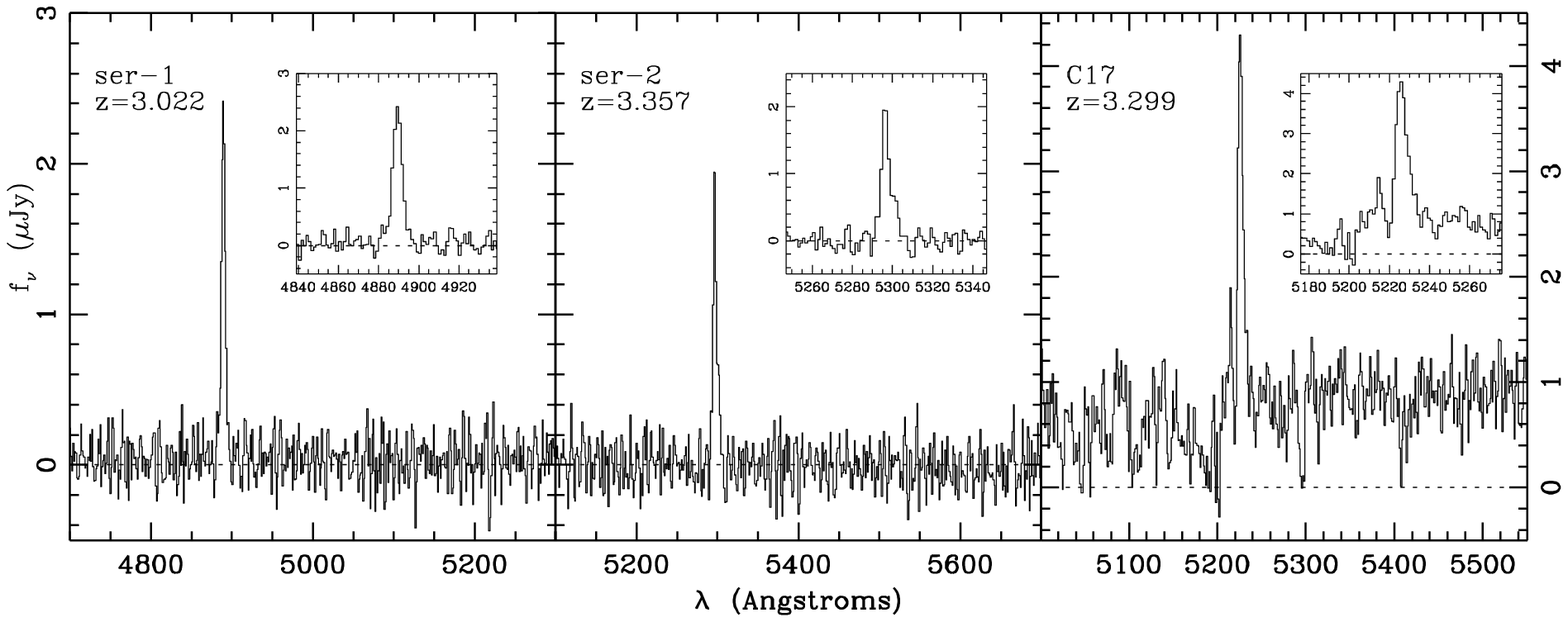}{2.0in}{0}{75}{75}{-235}{-369}

\caption{Spectra of two serendipitously identified \lya-emitters in the
field of SSA22, with the known LBG SSA22~C17.  The abscissa on the left
refers to the first two plots, while the abscissa on the right refers
to the right-hand plot.  Inserts illustrate the characteristic
asymmetric \lya\ profile of C17 and ser-2.}

\end{figure}

%[EDITOR: FIGURE 3 GOES NEAR HERE]

The spectra of the serendipitous emission lines ser-1 and ser-2 are
presented in Fig.~3, together with that of the LBG SSA22~C17.  The
redshifts of these galaxies, their emission line wavelengths, fluxes
and attributed luminosities are given in Table~1.  It should be noted
that the fluxes and luminosities represent only that part of the
atmospherically smeared image of the galaxy falling within the slitlet,
and are in all likelihood under-estimates as these
serendipitously-identified sources are not necessarily centered in the
slitlets.  The sources ser-1 and ser-2 are at redshifts $z = 3.022$,
and $z = 3.357$, respectively.  We note that ser-2 is at nearly the
same redshift as the nearby quasar SSA22~D14 \citep{Steidel:98}.

To date, our serendipitous search has reviewed five slitmasks ($\simeq
0.75$ arcmin$^{2}$) and five longslit spectra ($\simeq 0.30$
arcmin$^{2}$).  In addition to the two lines presented here, we have
found one other definite \lya\ line at $z = 2.946$ with a line flux
$j = 3.32 \times 10^{-17} ~\ergcm2s$ and line luminosity ${\cal
L}_{\rm Ly\alpha} = 2.2 \times 10^{42} h_{50}^{-2}\, \ergs$.  In a
total search area slightly over 1 arcmin$^{2}$, this suggests that the
surface density of robustly identified high equivalent width \lya\
emission lines within the redshift range, $2.5 \leq z \leq 3.5$ is
very roughly of the order $3 \pm 2$ arcmin$^{-2}$ per unit redshift.
This is consistent with the estimate of
\citet{Cowie:98} who find a surface density of $\approx 3.6$
\lya-emitters arcmin$^{-2}$ per unit redshift to a higher limiting
flux density of $2 \times 10^{-17} ~\ergcm2s$.  The full survey
including candidates at higher redshift will be discussed in a
subsequent paper.

\subsection{Intrinsic Angular Diameters and Velocity Dispersion}

We derive the intrinsic spatial FWHM assuming that the intrinsic,
atmospheric, and instrumental FWHM add in quadrature.  Based on on the
results of multiple measurements, we assume the 1$\sigma$
observational and intrinsic FWHM measurement errors to be $\sim 0.25$
pix, and run 100 trials based on the observed central values.  We find
deconvolved FWHM for ser-1 and ser-2 of 0\farcs24 $\pm$ 0\farcs18 and
0\farcs36 $\pm$ 0\farcs15, respectively.  At these redshifts, a
compact galaxy with $r=1 \, h_{50}^{-1}$ kpc would have an intrinsic
angular diameter of $\sim 0 \farcs 27 \, / \, 0 \farcs 16$ ($\Omega=1
\, / \, 0.1$).  We note, however, that the deconvolved spatial FWHM of
C17 at the \lya\ line, and in the continuum redward of the line are
$\sim 0 \farcs 90 \, {\rm and}\, 0 \farcs 40$, respectively.  The
difference is likely due to scattering of \lya\ photons in the \hone\
halo.  Hence it is conceivable that the intrinsic sizes of the stellar
components of ser-1 and ser-2 are $\sim 50 \%$ smaller than the values
obtained above, or $r \simlt 500 h_{50}^{-1}$ pc.

We measure the instrumental resolution (for objects filling the
slitlet) from the line profile of lamp spectra.  We find FWHM$_{\rm
instr}^{\rm lamp} \simeq 4.4$ \AA.  For spatially unresolved sources,
we find FWHM$_{\rm instr}^{\rm QSO} \simeq 2.8$ \AA, measured from the
\lya\ forest of the QSO SSA22~D14.  The deconvolved velocity width is
given by the equation, 
\begin{equation} \sigma_v = c \frac{\sqrt{({\rm
FWHM}_{\rm obs})^2 - ({\rm FWHM}_{\rm instr})^2}} {2.354\,
\lambda_{\rm obs}}, 
\end{equation} 
where $c$ is the velocity of light. We find that $\sigma_v$ is
consistent with $\sim 55 \pm 8.5 ~ \kms$ for both serendipitous
objects assuming the sources fill the slitlet (see Table 1).  For
unresolved sources, perhaps more appropriate for the small angular
extent of these sources, $\sigma_v \lesssim 95 ~\kms$.  Bear in mind,
however, that these values refer to the kinematics of the
circum-galactic H I cloud, rather than stellar velocity dispersion,
which will be less (see related discussion in \S5).

\subsection{\lya\ Equivalent Widths}

The measured continua for the serendipitously-identified \lya-emitters
were consistent with zero counts, confirming the visual impression of
ser-1 and ser-2 seen in Figs.~1 and 3.  Therefore, direct
spectroscopic measure of the line equivalent widths is not possible.
We can, however, determine lower limits on the equivalent widths using
the $1 \sigma$ uncertainty of the continua.  The Poisson noise is the
square root of the photon counts per pixel, and we denote its average
per pixel by $\sigma_{\rm pix}$.  The summed noise in a aperture of
width $w$ and length $l$ is $\sigma_{\rm pix} \sqrt{wl}$.  We have $l$
different approximations of the value of the continuum, so the noise
in the continuum $\sigma_{\rm cont}$ is, \begin{equation} \sigma_{\rm
cont} = \sigma_{\rm pix} \sqrt{\frac{w}{l}}.  \end{equation} However,
$w$ is not a free parameter, but must be chosen so as to maximize the
$S/N$.  Since our sources are small, we assume their undetected
continua have a Gaussian shape along a spatial cut, and that their
FWHM is the same as the seeing ($\sim$ 0\farcs78).  For the \lya\
line, this would imply a source of zero intrinsic extension.  But
since the FWHM of the continuum is probably significantly less than
that of the \lya\ line, as we found with SSA22~C17 (\S3.1), it may in
fact be a slight over-estimation of the continuum FWHM.  We find the
$S/N$ ratio is maximized when $w = 1.165 \times$ FWHM.  For $l$ we
chose 100 pixels ($\sim$ 126 \AA).  Then,
\begin{equation} W_{\lambda}^{\rm obs}\, (2 \sigma) \simgt
\frac{j_{\alpha}}{2 \sigma_{\rm cont}}.  
\end{equation} 
For the 1 $\sigma$ continuum uncertainties, $\sigma_{\rm cont} \simeq
1.7 (1.4) \times 10^{-20} ~\ergAcm2$ for ser-1 (ser-2).  We find the
$2 \sigma$ lower-limits on the observed equivalent widths are
$W_{\lambda}^{\rm obs} \geq 550$ \AA\ (ser-1), and $\geq 470$ \AA\
(ser-2).  The rest-frame values are tabulated in Table~1.  Considering
the apparent absence of \nv\ and \civ\, and the low ISM velocity
dispersion of these sources, the data is consistent with stellar
sources, and only marginally with a possible weak AGN.  We discuss
this further in \S4.  SSA22~C17 has measured $W_{\lambda}^{\rm obs}
\simeq 35.6$ \AA.

\subsection{Star Formation Rates}

Assuming the \lya\ line flux is powered by star formation, the
strength of this line can be used to infer a star formation rate for a
young galaxy.  We first estimate the H$\alpha$ luminosity
corresponding to the measured \lya\ luminsity: the low-density case B
ratio \lya/H$\beta$ is approximately 25 in the absence of dust
\citep{Ferland:85}, while H$\alpha$/H$\beta \sim 2.8$ for hot stars
\citep[$T \simeq 1.2 \times 10^4$~K;][]{Osterbrock:89}.
\citet{Kennicutt:83} provides the conversion from H$\alpha$ luminosity
to the SFR, resulting in a SFR -- ${\cal L}_{{\rm Ly}\alpha}$ relation
of $\dot{M}\, (\Msun$ yr$^{-1}) \simeq 10^{-42} {\cal L}_{{\rm
Ly}\alpha}$ (\ergs).  This results in SFRs of $\sim 1.3/1.2 ~\Msun$
yr$^{-1}$ for ser-1/ser-2, respectively.  Because of the possibility
of dust, or incomplete absorption of ionizing radiation by the gaseous
halo, these numbers are lower limits. We believe slit losses are
minimal as the lines are nearly unresolved and do not fill the slit.
Line fluxes from two other nights of observations support this
conclusion.  The continuum flux of the LBG SSA22~C17 at 1500 \AA\ was
used to determine the SFR of ${\dot M} = 27.7 ~\Msun$ yr$^{-1}$
\citep{Madau:98}, based on a Salpeter IMF with mass range $0.1 \leq
M/\Msun \leq 125$.  However, based on the synthetic spectra of
\citet{Leitherer:95}, with a continuous SFR at $\sim 10$ Myr of age,
the SFR is $\sim 1.6$ times larger, or $\sim 45 ~\Msun$ yr$^{-1}$.  We
expect that the true value is somewhere in between.  The \lya\ line
luminosity of C17 (Table 1) would imply a SFR of $2.2 ~\Msun$ yr.

\subsection{Projected Photometric Magnitudes}

We use the spectral slope of a $10^7$ year old starburst with a rest
equivalent width of 150~\AA \citep{Charlot:93} to calculate the
expected continua of ser-1 and ser-2.  We would thus expect continuum
magnitudes of $V=27.8$/28.0 for ser-1/ser-2, or $V=27.4$/27.7 when the
\lya\ line is in included.  The $2 \sigma$ upper limits on the
continua of ser-1 and ser-2 (\S3.2) are only $\sim 30 \%$ greater than
the continua implied by ${\rm W}_{\lambda}^{\rm rest}=150$ \AA.

\section{Discussion}

What is the physical origin of these isolated, high equivalent width
emission lines?  \nv\ or \civ\ are not found in the spectra of ser-1 or
ser-2.  The ratio of the high-ionization line flux relative to \lya\ is
$\simlt 1/15$ and $\simlt 1/20$ for \nv\ and \civ, respectively, significantly
less than ratios measured from composite AGN spectra
\citep[\eg~][]{Stern:99a}.  The small velocity dispersions of ser-1 and
ser-2 ($\sigma_v \simeq 60 \kms$) and the lack of associated
\nv\ or \civ\ emission argue against the presence of an AGN, while the
high surface brightness is inconsistent with photoionization by the
metagalactic flux \citep[\eg~][]{Bunker:98}.  Thus we assume the lines
owe their existence to stellar sources.  Below we argue that the
galaxies are primeval, and attempt to fit them into a cogent picture of
the evolving Universe at $z \simeq 3$.

\subsection{The Production of High Equivalent Width \lya\ Emission}

The production of a strong \lya\ line requires two things: first there
must be a large column of neutral hydrogen which can intercept almost
all photons with $\lambda \leq 912$ \AA, and second, \lya\ radiation
must escape the cloud.  The large cross section of \hone\ for \lya\
photons makes \lya\ extremely vulnerable to dust quenching.  We
consider the scenario of \citet{TenorioTagle:99} in some detail as one
model of the early phases of a starbursting protogalaxy.
\citet{TenorioTagle:99} suggest that after an initial stellar
ionization of dynamically quasi-relaxed gas and attendant high
equivalent width ``pure'' \lya\ emission (\ie\ a symmetric
line profile), enriched stellar winds sweep up gas.  The compressed
gas recombines, forming a shell of neutral gas which absorbs emerging
Lyman continuum radiation.  \lya\ radiation is then emitted and
scattered within the shell.  The expansion of the compressed shell of
gas results in the preferential scattering, and probable absorption,
of the blue-wing of the \lya\ line.  The LBG SSA22~C17 (Figs.~1, 3) is
a good example of this common feature of LBGs; in the insert we see a
trough with center $\sim 350 ~\kms$ blueward of the line center, while
at $\sim 650 ~\kms$ we see a second maximum.  Here, perhaps, is
palpable proof of a strong wind and a neutral shell, with the blueward
emission line probably caused by the recombination of shocked gas
associated with the expanding shell \citep{TenorioTagle:99}.  In this
scenario, the time scale for high equivalent width \lya\ line emission
must be only modestly greater than the lifetime of the most massive
stars (at most $\sim 10^7$ yr).

This model is based upon a single, central starburst driving the \lya\
emission --- an assumption which may not generally hold in real
galaxies.  High-resolution VLA/optical observations of five local
star-forming blue compact dwarfs reveal a history of spatially
distinct star forming regions in various stages of evolution
\citep{vanZee:98}.  Perhaps new star-forming regions in the same
galaxy could display high equivalent width \lya\ emission.  This
effect appears to be seen in a $z=4.04$ lensed galaxy whose star
formation regions are subjected to a detailed lensing chromatography
study by \citet{Bunker:00}.  Their color analysis shows a sequence of
non-coeval star formation sites with ages ranging from zero to $\sim
100$ Myr.  ``Nodes'' of star formation, showing \lya\ in absorption,
have spectral energy distributions consistent with a burst age of $20
- 30$ Myr.  There is, however, an outlying region of very high
equivalent width \lya\ emission ($W_{\lambda}^{\rm rest} > 100$ \AA,
$3 \sigma$) thought to be resonantly-scattered \lya\ photons escaping
from the adjacent star-forming node.  Yet summed over the whole galaxy
(as we would see it without the benefit of gravitational lensing), the
\lya\ equivalent width is modest.

We therefore suggest that, for apertures containing entire young
and compact galaxies, high equivalent width stellar \lya\ emission
($W_{\lambda}^{\rm rest} \simgt 100$ \AA) indicates a primeval or
near-primeval galaxy.  We conclude that it is very likely that ser-1
and ser-2 are primeval galaxies --- by which we mean they are
undergoing their first significant burst of star formation.

\subsection{The Surface Density of High-Redshift \lya-Emitters}

How do ser-1 and ser-2 fit within the context of surveys of star
forming galaxies at high redshift?  The luminosity function (LF) of
galaxies at high redshift is a useful comparative tool for our
purposes.  \citet{Steidel:99} find a steep faint-end slope ($\alpha
\simeq -1.6$) for the observed ${\cal R}$-band LF of star-forming
galaxies at $\langle z \rangle \simeq 3.04$, implying that a large
fraction of the UV luminosity density is produced by galaxies fainter
than the spectroscopic limits of current Earth-based photometric
surveys.  By virtue of the association of UV luminosity with SFR, this
implies that a large fraction of the star formation history is
unobserved.  When integrated to the apparent magnitudes expected of
the continua our serendipitous emission-line galaxies (${\cal
R} \simeq 28$ mag), their UV luminosity function implies number densities
$\sim 20$ times, and integrated star formation density $\sim 4$
times, higher than the classical LBGs integrated to the completeness
limits of their ground-based survey, ${\cal R}=25$ mag.

As we have noted, because the \lya\ emission lines of such sources as
our serendipitous objects are likely to be observable only for a very
small fraction of 1~Gyr, the modest numbers of high equivalent width
\lya\ sources are tangible evidence for this many-fold larger
population of slightly older, faint continuum, star forming galaxies
with self-absorbed \lya\, inaccessible to even the deepest
spectroscopic or color selection surveys.  If we are to believe that
high equivalent width \lya\ can only be exhibited for the first $\sim
5 \times 10^{7}$ yr of a galaxy's life, and that their surface density
is $\sim 3\ {\rm arcmin}^{-2}$ per unit redshift, then if their
distribution is uniform between $3 \leq z \leq 4$, there must be $\sim
30$ (50) galaxies ($\Omega=1~(0.1)$) per square arcminute formed in
that interval.  This follows because there are 10 ~(17) intervals of
$5 \times 10^7$ yr between $z=3$ and $z=4$.  The surface density of
newly formed galaxies would thus be $\sim 30$ times larger than the
LBG surface density.  One must integrate the UVLF to ${\cal R}=28.5$
to achieve that surface density.  Thus, available surface density
estimates of primeval (high equivalent width \lya\ emission) galaxies
are consistent with the claim that most galaxies at $z\simeq 3$ were
formed at redshifts less than $z=4$.  However, the integration of the
$z \sim 4$ LF \citep{Steidel:99} over luminosity limits identical to
that at $z=3$ shows that the implied co-moving number densities
increase only slightly for $\Omega$ in the range of $0.1 \leq \Omega
\leq 1.0$, apparently consistent with very low net galaxy formation.
These two lines of research could be consistent if the merger rate was
nearly equal to the galaxy formation rate.  There is evidence that the
epoch-dependent merger rate was higher in the past by $~(1+z)^m$ with
$m \sim 1.5-3.0$ \citep{Windhorst:99}, and that mergers were more frequent
for the sub-${\cal L}^*$ galaxies.  However, determining the
size-dependent merger rate that would be needed to leave the shape of
the luminosity function unchanged while producing only a very modest
increase in comoving galaxy density, is beyond the scope of this
paper.

\subsection{Environment}

The environment of high equivalent width \lya-emitters is of
considerable interest.  In their study of Lyman-break candidates in
the SSA22 field, \citet{Steidel:98} display convincing evidence for
redshift ``spikes'' at $\langle z \rangle = 3.09$, and $\langle z
\rangle = 3.35$.  Ser-1 lies $\sim 5000 \kms$ from the former
enhancement, a structure which they suggest is a proto-Abell cluster.
The latter redshift spike, which is substantially less massive than
the former, contains a quasar (SSA22~D14).  Ser-2 is only $\sim 190
(320) h_{50}^{-1}$ kpc (projected), for $\Omega = 1 ~(0.1)$, and
$\simlt 200 \kms$ distant from the quasar, while both are on order 400
\kms\ beyond the apparent center of the redshift spike\footnote{ We
modeled the effect of the quasar on ser-2 by extending the observed
continuum blueward of the \lya\ line.  The extrapolated photon flux is
such that its excitation of gas around ser-2 could account for at most
$\sim3 \%$ of the \lya\ emission of ser-2.}.  We note that the
serendipitously-identified high equivalent width \lya\ line, SSA22-S1
\citep{Steidel:98} at $z=3.100$, is $\sim 750 \kms$ from the $\langle
z \rangle = 3.09$ enhancement.  Thus, this limited amount of data
suggests that high equivalent width \lya\ emitters tend to be loosely
associated with density enhancements some of which may be
protoclusters.  Similar clustering of \lya\ emitters is seen in
studies at $z \simeq 2.4$ \citep{Keel:99}.

\section{A Primeval Galaxy Model}

In a speculative vein, we now consider what we have learned of
primeval galaxies in the light of observations, and the modeling of
associated phenomena.  Our purpose is to emphasize the significance of
the above results to the wider field of cosmology.  Our study suggests
the existence of a high volume density of tiny galaxies with high
specific SFRs, which may cumulatively have rivaled that of the LBGs.
We seek here to provide a galaxy formation scenario in which
this is plausible, and to place this scenario in the larger context of
the evolution of the universe.

In response to the very small effective radii of faint galaxies imaged
by HST \citep[\eg][]{Giavalisco:96b, Pascarelle:98}, it has been shown
that an ``inside-out'' galaxy formation scenario may produce galaxies
with small scale-lengths \citep{Bouwens:97, Cayon:96}.  A theoretical
framework for this hypothesis is presented in a paper by
\citet{Lin:92}, which seeks to model primeval galaxies using two
alternative hypotheses of protogalactic clouds.  In these models,
robust self--regulating star formation occurs at a rate inversely
proportional to the cooling time for hydrogen number densities greater
than $n_c \sim 4 \,{\rm cm}^{-3}$.  Their model $A$ is a homogeneous
cloud which collapses uniformly; the critical density is first reached
at a size $R_c \simeq 6 M_{11}^{1/3}$ kpc, where $M_{11}$ is the
galaxy mass in units of $10^{11} ~\Msun$, creating a star-forming
region whose size is inconsistent with observations of either ser-1 or
ser-2.  Model $B$ is an isothermal cloud with a core of $500$ pc,
which collapses in an inside--out manner according to the
\citet{Shu:77} self-similar collapse model, and reaches the density
$n_c$ at $R_c \simeq 1 M_{11}^{1/2} R_{100}^{-1/2}$ kpc , where
$R_{100}$ is the outer radius of the cloud in units of $100$ kpc.
Model B predicts sizes that are very close to the constrained sizes of
our serendipitous emission line regions, and makes their comparison
with the model appear quite promising.  It also predicts that
protogalaxies will have a relatively constant bolometric luminosity
for a period of $\sim 1.7$ Gyr.  In their published simulation, the
bolometric luminosity $\log{\cal L} \sim {43.5}\, (\rm{ erg \,
s}^{-1})$, though the scale-free nature of isothermal spheres could be
expected to allow the somewhat more modest luminosities observed here.
We shall see that this luminosity is consistent with a SFR of $\sim
5.7 ~\Msun \, {\rm yr}^{-1}$.

An isothermal cloud which is bounded by an external medium of pressure
$P_{\rm ext}$ may achieve a hydrostatic equilibrium if the ratio of
the external to the internal pressure is less than 14.3
\citep{Shu:77}.  This cloud, known as a Bonnor-Ebert sphere, has a
just-critical mass $M_{crit} \simeq 0.8 a^4 G^{-3/2} P_{\rm
ext}^{-1/2}$, where $a$ is the sound speed.  This mass may be
consistent with galactic masses when the intergalactic pressure is
modest (\ie\ $P_{\rm ext} \simlt 1 \times 10^{-19} {\rm dyne} ~{\rm
cm}^{-2}$).  Such a sphere would become super-critical, and collapse
from the inside out if it were part of a general over-density which
had pulled away from the Hubble flow and begun to collapse.  Linear
theory shows that the turn-around radius moves outward at a rate of
$\sim$half the Hubble flow \citep{Davis:80}, so that we could expect
the location of high equivalent width \lya\ emission to be somewhat
within the turn-around radius, and to move outward with time.  A more
extensive discussion of this subject will be reserved for a future
publication.

In order to link the model with observations, we made a rough
simulation of a galaxy undergoing a constant SFR over a period of up
to 1 Gyr.  We used a Salpeter IMF ($\alpha = -2.35$) with lower mass
limit of $0.1 ~\Msun$, and main sequence lifetimes and luminosities
(Tables 3-6 and 3-9) from \citet{Mihalas:81}, modeling them as black
bodies, which is sufficiently accurate for metal-free, non-evolved
stars.  Unless otherwise stated, our upper mass limit is $125 ~\Msun$,
and we assume that the SFR is $1.2~\Msun\,{\rm yr}^{-1}$.  The \lya\
photon emission rate of a primeval galaxy is taken to be 2/3 of the
stellar emission of photons more energetic than 13.6 eV, according to
the case B recombination \citep{Spitzer:78}.  We assume that $80\%$ of
all newly introduced gas (\ie\ over and above the $\sim 10^{7} ~\Msun$
of gas of density $n_c$ within the central 500 pc) is transformed into
stars immediately.

The continuum measured in the $75$ \AA\ region around the \lya\
emission line, together with the \lya\ luminosity, is used to
calculate the equivalent width.  We tested the derived equivalent
widths against the dust-free calculations of \citet{Charlot:93} for
continuous star formation and upper mass limits of 80 and $120
~\Msun$, and found values that were quite consistent with values
attained by interpolating between the bracketing Salpeter slopes which
they employed.  We find that for a 10 Myr old galaxy,
$W_{\lambda}^{\rm{rest}} = 160/200/250$\AA\ respectively for
80/100/120 $\Msun$ upper limits.  These derived values drop to $\sim
125/155/200$\AA\ by 100 Myr.  The relation of the resulting \lya\ line
luminosity to the SFR is in agreement with the case B predictions.
The model predicts a continuum in the observed visual band of $\sim
1.3 \times 10^{-20} ~\ergAcm2$ for a $z=3$ galaxy with ${\dot M}=1.2
~\Msun ~{\rm yr}^{-1}$.  This is within the $1 \sigma$ non-detection
limits of the continua of ser-1 and ser-2 ($1.7/1.4 \times 10^{-20}
~\ergAcm2$).

For an upper mass limit of 100 $\Msun$, and $\dot{M}=1.0~\Msun ~{\rm
yr}^{-1}$, the bolometric luminosity at 10 Myr is $\sim 5.5 \times
10^{42}~\ergs$.  Thus, the luminosity of the \citet{Lin:92} model B
corresponds to $\dot{M}\simeq 5.7~\Msun ~{\rm yr}^{-1}$.  We calculate
the mass to bolometric luminosity ratio of our model by adding the
mass in stars and dissipated gas, dividing it by the luminosity.  The
dark matter does not participate in the cloud's collapse since it is
not pressure supported.  It will be compressed somewhat by the
gradually increasing concentration of baryons in the center, however,
on the whole, the DM halo should have little effect on the velocity
dispersion of stars formed out of dissipated gas in the first few tens
of Myr since they are formed well within the core of the isothermal
sphere.  We find that the calculated ${\cal M}/{\cal L} \simeq 0.01$,
0.05, and 0.37 for $\log {t} = 7$, 8, and 9 (Gyr) respectively.  We
note that compact narrow emission line galaxies, which appear to be
likely lower-redshift counterparts, have mass-to-luminosity ratios,
${\cal M}/{\cal L} \approx 0.1$ \citep{Phillips:97}.

The expected stellar velocity dispersion, $\sigma_s = \sqrt{G
M_{stars+gas}/r}$, implies dispersions of model galaxies of from $\sim
14~/~ 37~/~$, and $110 ~\kms$ for $\log {t} =7$, 8, and 9 (yr),
assuming a radius of 500 pc.  The low $\sigma_s$ for the 10 Myr old
galaxy is actually reasonable, in view of the finding that the very
large cluster in the blue compact dwarf NGC 1705 \citep{Ho:96} has
$\sigma_s \sim 11\kms$.  A low $\sigma_s$ is also not in conflict with
our deconvolved velocity dispersions of $\sigma_v \sim 60 ~\kms$
(\S3.1) since the latter is a measure of both the turbulence of gas
(from infall and stellar winds) which give rise to the \lya\ resonance
line, as well as the possibly spatially resolved structure within the
slitlet.  On the basis of the observed $\sigma_v$ and deconvolved
radius $r$ of the LBG C17 (Table 1), we find that the galaxy mass is
$M_{\rm{LBG}} = 4.3 \times 10^9 ~\Msun$.  This galaxy would take
$\sim$1.2 Gyr to form at the SFR derived for ser-1 or ser-2.  However,
between $z=3.5$ and $z=3.0$, there are only $0.265/0.473$ Gyr for
$\Omega=1.0/0.1$.  Thus, LBGs would not be formed in a reasonable
amount of time from low-SFR objects if SFRs remain constant.  Below,
we consider the possibility that the SFR may increase with time

Our simple galaxy model has assumed that ${\dot M}$ is constant.
However, a realistic isothermal sphere will have a baryonic core,
assumed here to be $\sim 500$ pc.  In this case case the infall rate
will $not$ be constant in time, but given a sound speed of 10 $\kms$,
may increase during the first $\sim 5 \times 10^7$ yr.  In addition,
the negligible metallicity expected of primordial gas will make
cooling times, and the Jeans mass, larger\footnote{It has been
suggested \citep{Lin:92} that the IMF of primeval galaxies may be
heavily weighted in high-mass stars, which could result in rest \lya\
equivalent widths significantly larger than $150$~\AA.  The truth of
this contention could be tested if spectroscopic observations deep
enough to detect the continua were made.} during the first few tens of
Myr of a galaxy's life.  Thus the effect of metals ejected from
evolved stars in aiding cooling, and of dust in catalyzing the
production of molecules, is to compound the effect of increasing
infall; the SFR and its efficiency can be expected to rise
substantially by a few tens of Myr.

On the other hand, recall that the high equivalent width of the \lya\
emission line of primeval galaxies is expected to be quenched over
essentially the same time scale.  This means that the nature, and fate
of these pre-LBG/post-primeval galaxies will be very difficult to
learn.  With an ernhanced SFR, the post-primeval galaxies could
plausibly grow by a factor of 10-40 into LBGs in a few tenths of a Gyr
or, by virtue of competition from neighbors, have their growth
truncated while still in the dwarf stage, or even be cannibalized;
environment may have a dominant role to play.

While the mass and SFRs of our purportedly primeval galaxies are much
smaller than that of LBGs, the $specific$ SFR of the former is on
average 10 or 20 times greater than the latter.  The large numbers of
small, young galaxies which are thought to have been formed (\S4.2) in
the interval $3 \simlt z \simlt 4$ might thus be expected to provide a
substantial fraction of the star formation rate density.  In fact, the
steep faint end slope of the $z \sim 3$ UV luminosity function of
\citet{Steidel:99} implies that up to 40\% of the SFR density may be
supplied by galaxies fainter than ${\cal R} =27$ (${\cal L} \sim 0.1$)
but brighter than the projected continuum magnitudes of our
serendipitous sources.  

In addition to the formation of a galactic bulge according to the
Lin-Murray process, it is expected to be accompanied by the formation
of the halo, as infalling clumps of gas dissipatively interact with
the already accreted gas \citep{Binney:76, Manning:99a}.

\section{Conclusions}

We draw the following conclusions from this study:

\begin{itemize}

\item We suggest that the combination of high equivalent widths, low
velocity dispersions, and small intrinsic sizes seen in the \lya\
lines of these two galaxies can be well-explained as their being
extremely low-metallicity galaxies undergoing an initial burst of star
formation--- hence, primeval galaxies.  These characteristics,
markedly different from earlier expectations, suggests an inside-out,
rather than a monolithic collapse formation mechanism.

\item Our data, which imply that most galaxies at $z=3$ may have been
formed at $z \leq 4$, stand in apparent conflict with the $z=3$ and 4
UVLFs \citep{Steidel:99}, which suggest that the comoving galactic
density remained relatively constant during this interval. These may
be reconciled, however, with a large merger rate that preserves the
faint-end slope of the luminosity function.

\item The agreement of the angular size, and the projected total
luminosity of these emission line galaxies, with the predictions of
the inside-out collapse of a nearly-isothermal \hone\ halo
\citep{Lin:92} suggests that the latter is a promising basis for
modeling galaxy formation.  Our discussion has suggested that a rising
star formation rate with a time-scale of $\sim 5 \times 10^7$ yr is
plausible for many post-primeval galaxies.

\item The isothermal collapse model requires that protogalactic clouds
form in relatively isolated regions.  However, their collapse may be
stimulated by the gradually increasing pressures of a protocluster
environment, thus accounting for the apparent weak clustering on large
scales noted of high-, as well as low-redshift \lya\ emission line
galaxies.

\end{itemize}

\acknowledgments

We thank C. Leitherer, and C. C. Steidel, for useful discussions.  We
are grateful to our referee, Dr. R. A. Windhorst, for valuable
suggestions.  We also thank A. Dey for assistance with the
observations.  We are grateful for the support of NSF grant AST
95-28536.

\bibliographystyle{apj}
% give path to manning.bib file as 2nd argument below.  ie,
% if manning.bib is at /moscow/kremlin/manning.bib, the
% argument should be /moscow/kremlin/manning
% you might use:
%\bibliography{apj-jour,/chakra/manning/serendip/SERsearch/dans_version/manning}
% my version uses:
%bibliography{apj-jour,/bigz1/manning/manning}
\bibliography{apj-jour,/chakra/manning/serendip/SERsearch/apj_replysubmit/manning}

\end{document}